# Modeling and Simulating Network Connectivity in Routing Protocols for MANETs and VANETs

W. Arshad[1], N. Javaid[1], R. D. Khan[2], M. Ilahi[1], U. Qasim[3], Z. A. Khan[4]

[1,2]COMSATS Institute of Information Technology, [1]Islamabad, [2]Wah Cant, Pakistan.
[3]University of Alberta, Alberta, Canada.
[4]Faculty of Engineering, Dalhousie University, Halifax, Canada.

**ABSTRACT**
This paper presents a framework for node distribution with respect to density, network connectivity and communication time. Using NS2, we evaluate and compare performance of three routing protocols; Ad-hoc On-demand Distance Vector (AODV), Dynamic Source Routing (DSR) and Fisheye State Routing (FSR) both in MANETs (IEEE 802.11) and VANETs (IEEE 802.11p). We further enhanced these protocols by changing their routing information exchange intervals; MOD AODV, MOD DSR and MOD FSR. A comprehensive simulation work is performed for the comparison of these routing protocols for varying motilities and scalabilities of nodes. As a result, we can say that AODV outperforms DSR and FSR both in MANETs and VANETs.

## I. INTRODUCTION

Mobile Ad-hoc Network (MANETs) are self-configuring network of mobile nodes connected by wireless links. Vehicular Ad-hoc Networks (VANETs) belong to special class of MANETs. They are distributed and self-assembling communication networks that are made up of multiple autonomous moving vehicles and peculiarize by very high node mobility.

Routing protocols are designed to calculate paths for communication networks. In table driven, proactive protocols are based upon periodic exchange of control messages and maintains routing tables. However, the reactive protocol tries to discover a route only when demand arrives. Our simulation work based upon three protocols comparison in MANETs and in VANETs named as AODV [1], DSR [2] and FSR [3]. Moreover, we introduce some modifications in their routing exchange intervals; 1) in MOD AODV, augments AODV's Expanding Ring Search algorithm (ERS) limits, 2) in MOD DSR, time associated with storage of routes in Route Cache is modified and 3) Scope intervals in FSR are adjusted in MOD FSR.

## II. Related Work and Motivation

In [4] [5], communication time between nodes is found when the nodes are moving in same and opposite direction with same or different speeds. In our work we improvement the work of [6] [7] and calculate the probability of link establishment between nodes when they are moving in same and opposite direction with same or different speeds.

The study [6] involved the consistently varying network topology and comparison of DSR with AODV in MANETs for different scenarios and performance metrics to propose the best scenario for each routing protocol to maximize its efficiency. In [7, 8] the authors modified the OLSR protocols in their paper. We also do some modifications and evaluate AODV, DSR and FSR for both MANETs and VANETs.

## III. Modeled Mathematical Framework

This work determine the steady-state distribution of the number of nodes within each segment. Let $N_{kj}$ is Poisson distribution with the parameter $\tilde{\lambda}_{kj} \int_0^t B_{x(\tau)}(R_{kj}) d\tau$. The probability distribution of the number of the nodes within segment $i$ and its probability generating function (PGF) at the steady state is given by:

---

$$Pr(\tilde{N}_i = n) = e^{-\phi_i} \frac{\phi_i^n}{n!} \quad and \quad P(z) = E[z^{N_i}] = e^{-\phi_i(1-z)} \quad (1)$$

Fig.1 shows detail scenario of communication time between node with probability of link establishment.

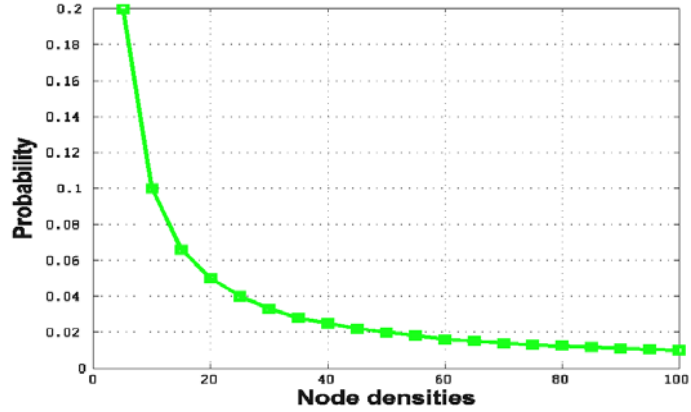

Figure 2: Probability of Communication

## IV. SIMULATIONS AND DISCUSSIONS

In this paper, simulations are performed on two Mac layer protocols; 802.11 for MANETs and 802.11p for VANETs. We evaluate and compare the performance of three selected routing protocols by their default values; AODV, DSR and FSR and modified values; MOD AODV, MOD DSR and MOD FSR, with different scalabilities and varying mobilities in MANETs as well as in VANETs. we modify these chosen protocols and then analyze their results.

For MOD AODV, changes have been made in default AODV's ERS algorithm; $TTL\_START = 2$, $TTL\_INCREMENT = 4$ and $TTL\_THRESHOLD = 9$. In MOD DSR, the modification made to original DSR consisted of reducing the size of Route Cache as taking $TAP\_CACHE\_SIZE = 256$. Smaller Route Cache means that only relatively fresh routes are stored. In MOD FSR, intervals of inner and outer scopes of FSR are changed to $1s$ and $3s$ respectively.

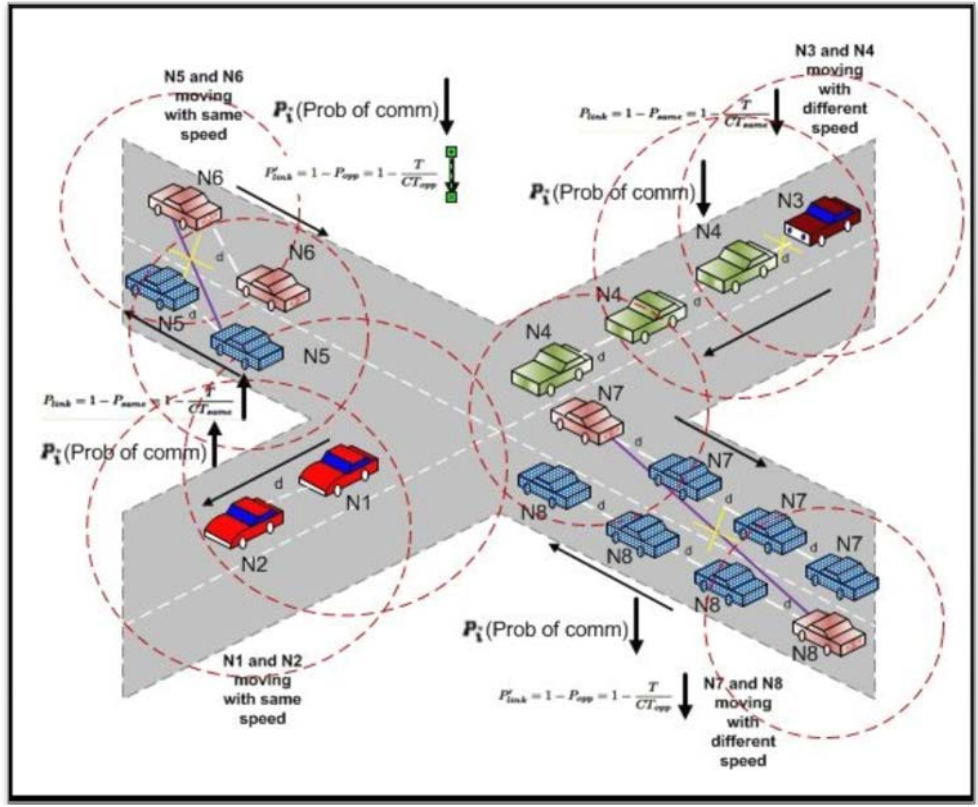

Figure 1: System Model

***Throughput*** Fig. 3.a,c shows that in smaller scalabilities, DSR has highest throughput and FSR has the lowest throughput. This is because of route caching mechanism of DSR, whereas the scope routing of FSR is best suited for very large networks consisting of thousands of nodes. On the other hand, in higher scalabilities, AODV and MOD AODV both perform best and DSR has lowest throughput in MANETs. AODV provides more communication time as specified in equation because of the local link repair.

In simulation results, Fig. 3.a,c depicts that MOD DSR gives better results in MANETs and in VANETs. We change cache size by decreasing its value to one fourth of its default value. By modifying $TAP\_CACHE\_SIZE$, fresher routes are available in the cache. As there is no explicit mechanism to delete stale routes in DSR, this modification helps to provide accurate routes as in equation (Probability of links due to fresher routes result more throughput).

In MANETs, MOD AODV not only improves its efficiency as compared to AODV but also outperforms among all other protocols. There is a less expanding ring values in initial default ERS values up to $TTL\_THRESHOLD$. We expand these rings by increment the $TTL\_INCREMENT$ value from $2$ to $4$. It lessens routing delay and increase communication probability as from equation. Also this results, not only low routing load but also lowers the routing latency. Ultimately, the throughput value increased.

Table 1: Simulation Parameters

| Parameters | Values |
|---|---|
| Simulator | NS-2(Version 2.34) |
| Channel type | Wireless |
| Radio-propagation model | Nakagami |
| Network interface type | Phy/WirelessPhy, Phy/WirelessPhyExt |
| MAC Type | Mac /802.11, Mac/802.11p |
| Interface queue Type | Queue/DropTail/PriQueue |
| Bandwidth | 2Mb |
| Packet size | 512B |
| Packet interval | 0.03s |
| Number of mobile node | 25 nodes, 50 nodes, 75 nodes,100 nodes |

| Speed | 2 m/s,7 m/s,15 m/s,30 m/s |
|---|---|
| Traffic Type | UDP, CBR |
| Simulation Time | 900 s |
| Routing Protocols | AODV, DSR, FSR, MOD AODV |
|  | MOD DSR, MOD FSR |

We observe that for low mobilites in MANETs MOD DSR outperform all other routing protocols; because DSR uses stale routes in frequent varying network topologies, on the other hand MOD DSR delete stale routes frequently as compared to DSR. FSR due to absence of trigger updates performs worst among all. In high mobilities, link breakage is frequent, packet salvaging (PS) only efficient in moderate and in low dynamicity. To maintain the fresher routes route cache updating time interval must be shorten to avoid stale routes. After shortening the cache size, its efficiency becomes almost equal to original AODV in VANETs, as shown in Fig. 3.d. Grat. RREPs help DSR and AODV to have higher throughput in mobility scenario. If we evaluate overall performance of all protocols in both MANETs and in VANETs then AODV which produces highest throughput, because it uses LLR, HELLO messages and gratuitous RREPs its advantage in highly mobile scenarios as depicted from Fig. 3.b,d.

*E2ED* It is the time required for a packet to reach its destination. In MANETs for low scalabilities, DSR produce highest E2ED because of first checking of route cache in low densities; the chance of alternate routes in route cache is less as compared to more number of nodes, thus produce more delay. MOD DSR, AODV, MOD AODV, FSR and MOD FSR have E2ED in descending order respectively. FSR and MOD FSR possess the lowest delay in both MANETs and VANETs. Proactive protocol gives minimum value of E2ED as compared to reactive nature protocols, from Fig. 4.a. Furthermore, the periodic intervals (inter and intra scopes) in MOD FSR is lower than FSR, that is why MOD FSR maintains relatively updated view of network topology which means fresher routes are more frequently updated. In VANETs as shown in Fig. 4.c the values of E2ED are much less than those in MANETs but the sequence remains the same.

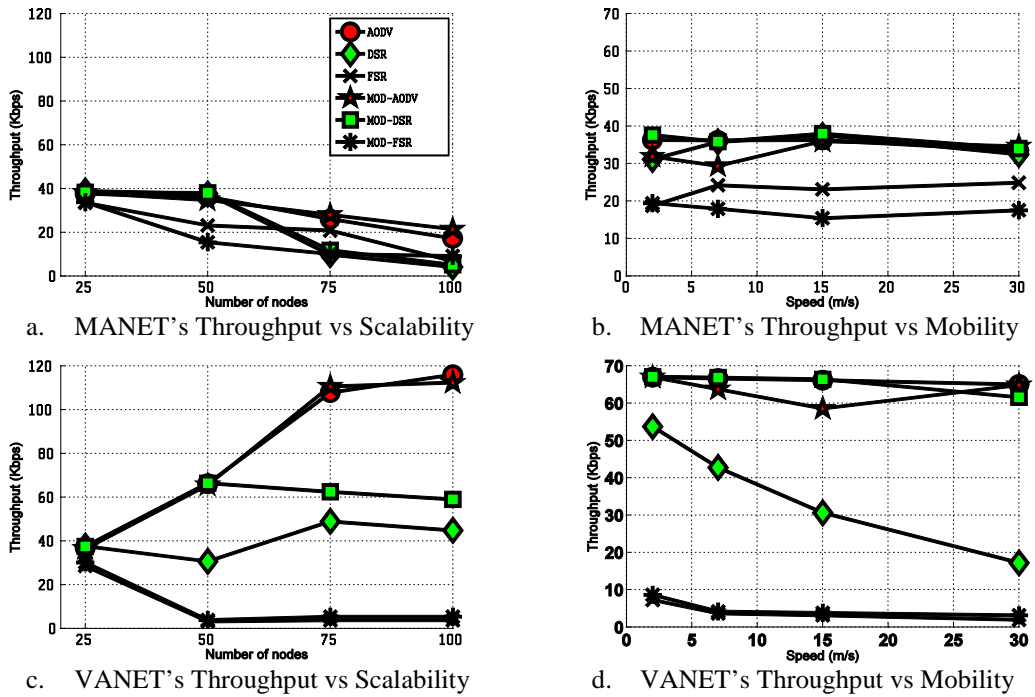

a. MANET's Throughput vs Scalability
b. MANET's Throughput vs Mobility
c. VANET's Throughput vs Scalability
d. VANET's Throughput vs Mobility

Figure 3: Throughtput

In lower mobilities DSR and MOD DSR has lower E2ED than AODV and MOD AODV, because of packet salvaging of DSR reduces the delay. While MOD FSR is the one with lowest E2ED due to its proactive nature and its frequent periodic updates. MOD DSR because of capability of storing fresher routes in route cache gives slightly low E2ED as compared to DSR. MOD FSR gives lower E2ED in both VANETs and MANETs as shown in the Fig. 4.b,d DSR uses route cache mechanism which works really well for less and moderate dynamic topologies. While AODV store only one rout for one destination for small interval. Therefore AODV has to find a new route to destination more often than DSR. The reduced cache size of MOD DSR enables source node to search quickly through route cache; which saves time consumption. Additionally, less number of faulty routes is stored because MOD DSR only stores fresh routs as compared to DSR. The route caching mechanism of DSR fails (sort of) in high mobility, where as AODV has ERS mechanism with LLR without caching, thus introduce more delay.

*NRL* It is the number of control messages transmitted to receive one data packet. In scalabilities scenario with MANETs, DSR attains lowest NRL in low scalabilities, on the other hand in high scalabilities generating the highest NRL grat. RREPs generated by intermediate nodes in low scalabilities are more suitable, while these RREPs in high densities producing high routing load. MOD DSR, produces slightly less NRL than DSR, because of fresher routes in route cache provides accurate information, thus avoiding route rediscovery. AODV produces more routing load as compared to MOD AODV, as ERS value is more in MOD AODV and make it more scalable. More frequent exchange of routing updates in MOD FSR generates more NRL than FSR, as shown in Fig. 5.a. In VANETs AODV, MOD FSR, MOD AODV, FSR, MOD DSR and DSR generated NRL in descending order as mention in Fig.5.c. AODV not only uses grat. RREPs during route discovery but also *HELLO messages* and LLR mechanism during route maintenance, thus overall it attains the highest NRL.

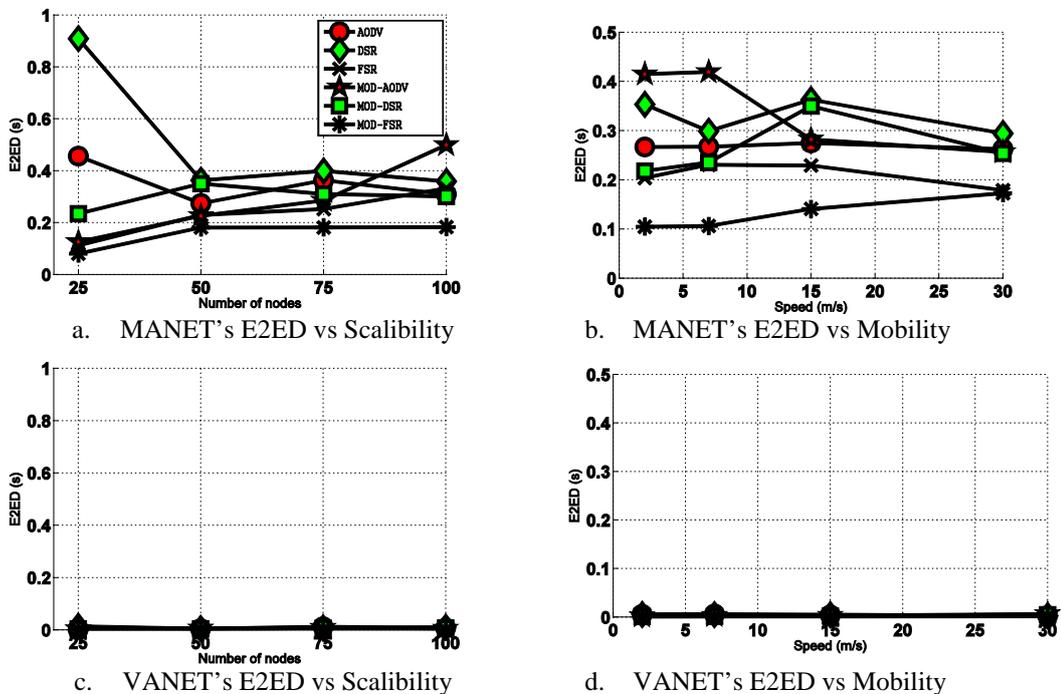

a. MANET's E2ED vs Scalibility  
b. MANET's E2ED vs Mobility  
c. VANET's E2ED vs Scalibility  
d. VANET's E2ED vs Mobility  

Figure 4: E2ED

DSR generate high NRL in selected mobilities for MANETs because of gratuitous RREPs and packet salvaging. Generally, packet salvaging and grat. RREPs are used to reduced routing load, but in high densities stale routes causes more routing overhead. MOD DSR possesses low routing load as compared to DSR, because of frequently updation of route cache. FSR generated the lower NRL because of graded frequency techniques as well as having scope view of topology. MOD FSR generates higher NRL because of small periodic intervals. The Medium Access Control protocol in IEEE 802.11p uses the Enhanced Distributed Channel Access (EDCA) mechanism originally provided by IEEE 802.11e as mentioned in [9] about MANETs and VANETs. DSR is a protocol which depends upon the link sensing on MAC protocol, therefore in scalabilities NRL varies in both MANETs and VANETs.

In mobility scenario MOD FSR has maximum NRL while AODV, MOD AODV also produce high NRL. FSR and MOD FSR produced high NRL because of proactive nature. As number of nodes are constant and mobilities varies. AODV LLR produces high NRL because LLR is best suited for denser networks. MOD AODV produces less NRL then AODV due to less repetitive RREQs bcause of less number of rings as compared to original AODV. DSR and MOD DSR produce low NRL respectively as shown in Fig. 5.b. DSR produces small amount of NRL because of route caching. In MOD DSR the NRL is even lower because of smaller size of rout cache reduce packet salvaging due to incorrect routes in route cache. In VANETs, sequence remains the same shown in Fig. 5.d. All the results taken for NRL shows that minimum routing load is provided by DSR. The increasing number of sources does affect NRL and it may behave different than expected in some situations. In Fig. 5.b,d reasonable as AODV generates most of its NRL due to grat. RREPs, LLR and $HELLO\ messages$.

## V. CONCLUSION AND FUTURE WORK

In this paper, a framework is presented for node distribution with respect to density, network connectivity and communication time. Routing protocols DSR, AODV and FSR were compared in MANETs and VANETs. Besides evaluating we also made some modifications to these routing protocols and observed their performance. These changes provide better result. We conclude that AODV performs best among original protocols while MOD DSR produces highest throughput. In high speeds, DSR due to stale routes in route cache fails to converge. On the other hand in MOD DSR due to reduction in the size of route cache improves overall performance in high speeds.

In future, we are interested to apply the same analysis on quality link metrics proposed in [10-12] and at MAC layer as [13, 14].

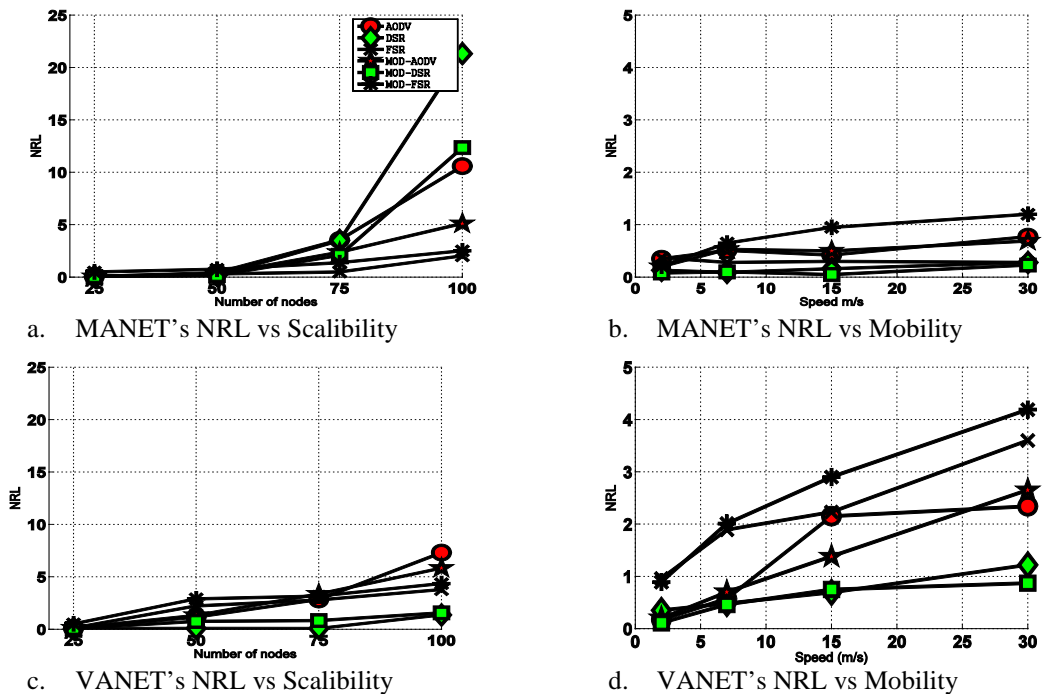

a. MANET's NRL vs Scalibility
b. MANET's NRL vs Mobility
c. VANET's NRL vs Scalibility
d. VANET's NRL vs Mobility

Figure 5: NRL

## REFERENCES

[1] C. Perkins, et al. IETF RFC3561, "Ad hoc On-Demand Distance Vector (AODV) Routing". Available: http://www.ietf.org/rfc/rfc3561.txt

[2] D. Johnson, et al., "DSR: The Dynamic Source Routing Protocol for Multi-Hop Wire-less Ad Hoc Networks," in In Ad Hoc Networking, 2001.

[3] G. Pei, et al., "Fisheye State Routing in Mobile Ad Hoc Networks," in ICDCS Work- shop on Wireless Networks and Mobile Computing, 2000.

[4] C. Gerald and P.Wheatley, "Applied Numerical Analysis", Addison-Wesley Publishing Company, 1994.

[5] Saad M. Almajnooni and Basem Y. Alkazemi, "Route Breakage Modeling of Mobile Ad Hoc Networks in Motorway Scenario", International Conference on Wireless and Mobile Communications, 2010.

[6] Samir R. Das, et al., "Performance Comparison of Two On-demand Routing Protocols for Ad Hoc Networks".
[7] Imran Khan "A Performance Evaluation Of Ad Hoc Routing Protocols for Vehicular Ad Hoc Networks" Thesis Presented to Mohammad Ali Jinnah University, 2009.
[8] M. Khabazian and M. K. Mehmet Ali, "A Performance Modeling of Vehicular Ad Hoc Networks (VANETs)", IEEE Communication Society WCNC, 2007.
[9] Bhavyesh Divecha, Ajith Abraham, Crina Grosan, and Sugata Sanyal "Impact of Node Mobility on MANET Routing Protocols Models".
[9] Usman A. etal., "An Interference and Link-Quality Aware Routing Metric for Wireless Mesh Networks,". IEEE 68th Vehicular Technology Conference, 2008.
[10] Javaid. N, Javaid. A, Khan. I. A, Djouani. K, "Performance study of ETX based wireless routing metrics," 2nd IEEE International Conference on Computer, Control and Communications (IC4-2009), Karachi, Pakistan, pp.1-7, 2009.
[11] Javaid. N, Bibi, A, Djouani, K., "Interference and bandwidth adjusted ETX in wireless multi-hop networks", IEEE International Workshop on Towards Samart Communications and Network Technologies applied on Autonomous Systems (SaCoNaS2010) in conjunction with 53rd IEEE International Conference on Communications (ICC2010), Ottawa, Canada, 2010., Miami, USA, 1638-1643, 2010.
[12] Javaid. N, Ullah, M, Djouani, K., "Identifying Design Requirements for Wireless Routing Link Met- rics", 54th IEEE International Conference on Global Communications (Globecom1012), Houston, USA, pp.1-5, 2011.
[13] Dridi. K, Javaid. N, Daachi. B, Djouani. K, "IEEE 802.11 e-EDCF evaluation through MAC-layer metrics over QoS-aware mobility constraints", 7th International Conference on Advances in Mobile Computing & Multimedia (MoMM2009), Kuala Lumpur, Malaysia, 2009
[14] Dridi. K, Javaid. N, Djouani. K, Daachi. B, "Performance Study of IEEE802.11e QoS in EDCF-Contention-Based Static and Dynamic Scenarios", 16th IEEE International Conference on Electronics, Circuits, and Systems (ICECS2009), Hammamet, Tunisia, 2009.